\documentclass[sigconf]{acmart}
\usepackage{booktabs} % For formal tables

\begin{document}
%\copyrightyear{2018} 
%\acmYear{2018}
%\setcopyright{acmcopyright} 

\acmConference[CryBlock'18]{1st Workshop on Cryptocurrencies and Blockchains for Distributed Systems }{June 15, 2018}{Munich, Germany}
\acmBooktitle{CryBlock'18: 1st Workshop on Cryptocurrencies and Blockchains for Distributed Systems , June 15, 2018, Munich, Germany}

\acmPrice{15.00}
\acmDOI{10.1145/3211933.3211947}
\acmISBN{978-1-4503-5838-5/18/06}

% --- End of Author Metadata ---

\title{From Bitcoin to Bitcoin Cash: a network analysis}
%

%
% 1st. author
%\alignauthor
\author{Marco Alberto Javarone}
\affiliation{%
  \institution{nChain}
  %\streetaddress{London W1W 8AP, UK}
  \city{London}
  \state{UK}
  %\postcode{W1W 8AP}
}
\affiliation{%
  \institution{School of Computing, University of Kent}
  \city{Medway}
  \state{UK}
}
\email{marcojavarone@gmail.com}
  
% 2nd. author
%\alignauthor
\author{Craig Steven Wright}
\affiliation{%
  \institution{nChain}
  %\streetaddress{London W1W 8AP, UK}
  \city{London}
  \state{UK}
  %\postcode{W1W 8AP}
}
\email{craig@ncrypt.com}

\date{04 May 2018}

\begin{abstract}
Bitcoins and Blockchain technologies are attracting the attention of different scientific communities. In addition, their widespread industrial applications and the continuous introduction of cryptocurrencies are also stimulating the attention of the public opinion.
The underlying structure of these technologies constitutes one of their core concepts. In particular, they are based on peer-to-peer networks.
Accordingly, all nodes lie at the same level, so that there is no place for privileged actors as, for instance, banking institutions in classical financial networks.
In this work, we perform a preliminary investigation on two kinds of network, i.e. the Bitcoin network and the Bitcoin Cash network.
Notably, we analyze their global structure and we try to evaluate if they are provided with a small-world behavior.
Results suggest that the principle known as 'fittest-gets-richer', combined with a continuous increasing of connections, might constitute the mechanism leading these networks to reach their current structure. Moreover, further observations open the way to new investigations into this direction.
\end{abstract}

\keywords{Bitcoin, Bitcoin Cash, Complex Networks}

\maketitle

\section{Introduction}
Nowadays, a number of services and platforms are based on distributed networks. In particular, one of the major benefits of distributed networks is given by the partition of a computational workload among multiple nodes, so that each one can perform an autonomous processing. As result, at a global level, a distributed network allows to implement the so-called 'parallel computing'~\cite{parallel01}.
When this kind of network is not controlled by a (or a few) central unit (e.g. a node that coordinates the whole system, or a part of the network), it can be also defined as 'decentralized'.
Blockchain is a modern technology, described for the first time in~\cite{satoshi01}, that can be briefly defined as a decentralized and distributed ledger. The latter contains records of transactions and implements different functionalities mostly based on the modern cryptography. 
In the last years, this technology is finding application in several industrial sectors~\cite{block01}, spanning from finance to healthcare, thus having an impact at different societal levels.
In this context, bitcoins are exchanged among users and the transactions are recorded in the Blockchain. In addition, as observed during last months, bitcoins are exponentially increasing their value, and a number of new cryptocurrencies is continuously generated~\cite{barronchelli01}.
The absence of a central control unit (e.g. a Banking Institution) appears to be one of the major and most notable results of Blockchain but, at the same time, can constitute a motivation of concern and skepticism for those who do not understand the underlying mechanism.
Since this technology is strictly based on a networked system, as older sharing platforms (e.g. those commonly used for file sharing), it is worth to investigate the related topological properties. Notably, a global topological view allows both to obtain a deeper knowledge on the structure, and to evaluate the dynamics of stochastic processes, as spreading and percolation, that can be of interest for new applications, security reasons, and so on.
Therefore, in this work, we aim to analyze the structure of these networks, with a focus on Bitcoin and Bitcoin Cash. Notably, the latter results from the former, and constitutes a new cryptocurrency (available in the market from August 2017).
Our results indicate that these two networks share some topological similarities, and that some generative models as 'preferential attachment'~\cite{barabasi01} or 'fittest-gets-richer'~\cite{bianconi01,javarone01} might be adopted for representing their evolution. 
In particular, even if both networks are 'peer-to-peer', parameters like the 'fitness' can be useful for discriminating nodes, e.g. from those provided with high computational resources to those with low power (e.g. Raspberry pi).
The remainder the paper is organized as follows: Section~\ref{sec:bitcoin_network} briefly summarizes the dynamics of the Bitcoin Network. Section~\ref{sec:network_analysis} introduces some concepts of complex network analysis and illustrates two generative models that might be useful for the considered systems. Section~\ref{sec:result} shows results of the numerical investigations. Finally, \ref{sec:conclusion} ends the paper.
\section{The Bitcoin Network}\label{sec:bitcoin_network}
In this section, we provide a very brief introduction to the Bitcoin Network~\cite{antinopolis01}. In particular, we aim to link the basic mechanisms of this network, without to consider all the local processes that occur during a single transaction, with the emergence of a connectivity pattern in the real dataset (below described) we consider in this analysis.
In few words, the Bitcoin Network represents the set of nodes running the bitcoin P2P protocol. Here, each node can have a specific role depending on its functionality, e.g. routing, mining, Blockchain database and so on.
Usually, a node that performs all these functions is defined as 'full node'. All nodes play as router in order to simplify spreading processes over the network.
Full nodes can verify any transaction without asking for external references. Mining nodes take part to a kind of 'competition' for solving the proof-of-work algorithm. Notably, this kind of nodes can also cluster together, forming 'mining pools', in order to increase their success probability (i.e. in the mentioned competition).
Now, when a new node joins the network, it has to generate connections with some of the pre-existing nodes. In particular, the new node has to discover at least one node in the network. This process is completely random, i.e. the new node connects with a pre-existing one randomly chosen. It is worth to remark that the Client used to join the network contains the list of some nodes (i.e. seed nodes). However the first nodes to consider for generating the new connections can be also those connected to seed nodes.
Finally, in order to be connected with the network in a reliable way, the new node generates a few connections, forming then different paths. Notably, while one single connection might be too little for ensuring reliability, usually nodes need only few neighbors, saving network resources.
\section{Network Analysis}\label{sec:network_analysis}
As stated before, in this context we are dealing with a networked system. Thus, the modern mathematical framework that allows to analyze this kind of systems is the Theory of Complex Networks~\cite{estrada01}. 
Due to its relevance and ubiquity in a number of fields, we provide a quick introduction. Notably, we focus on global properties for characterizing the structure of a network and on few generative mechanisms that can be of interest for studying the Bitcoin Network. 
Let us begin introducing simple concepts. A complex network is a graph characterized by non-trivial topological features. In more general terms, a graph is a mathematical entity that allows to represent relations among a collection of objects. More formally, a graph $G$ is described as $G=(N,E)$, with $N$ set of nodes and $E$ set of edges (or links). Nodes are the main elements of a system and can be described by a label. On the other hand, the edges represent connections among nodes and can map specific relations, e.g. friendship in a social network, correlations in EEG networks, direct links among websites in the WEB, and so on and so forth.
A graph can be `directed' or `undirected', i.e. we can have symmetrical relations among nodes or not. A simple line between two nodes represents an 'undirected edge', for instance to represent symmetric relations like friendship. Instead, in the second case, an arrow can represent a 'directed edge', for instance the link between two websites.
In addition, a graph can be `weighted' or `unweighted'. Notably, if a numerical value is associated to edges, for example because the related relations can be somehow weighted (e.g. representing an intensity), the graph is weighted. For instance, considering an airline network where the airports are mapped to nodes and the routes to edges, weights can be computed according to the geographical distance between pairs of airports.
The whole set of edges, of a $N$ nodes graph, is collected in a $N \times N$ matrix, defined `adjacency matrix'. Undirected graphs have a symmetric adjacency matrix, and unweighted graphs are represented by a binary adjacency matrix. In particular, the adjacency matrix $A$ of an unweighted graph is composed by the elements:

\begin{equation}\label{eq:adjacency_matrix}
a_{ij} = 
\begin{cases}
1 & \mbox{if $e_{ij}$ is defined}\\ 
0 & \mbox{if $e_{ij}$ is not defined}
\end{cases}
\end{equation}
Instead, a weighted graph is represented by a real matrix.
Now, we present with more details some global properties, of a general network, that can be computed by analyzing the adjacency matrix, i.e. the degree distribution, the clustering coefficient, and the path length.
\subsection{Degree distribution}
Nodes of a graph can have many connections (i.e. many edges). Usually, the number of connections of a node is called degree, and it is denoted as $k$. An important property widely used to asses the structure of a network is the degree distribution $P(k)$~\cite{barabasi01}. The latter represents the probability that a randomly selected node has the degree equal to $k$, i.e. connected with $k$ nodes.
Remarkably, the degree distribution uncovers a number of information related to a network. 
In addition, there are classes of random networks, as Erd\"{o}s-Renyi graphs and Scale-free networks, whose structure is well described in their degree distribution.
Early approaches to (complex) random networks have been developed by Paul Erd\"{o}s and Alfred Renyi~\cite{erdos01}. Notably, they defined a famous model known as Erd\"{o}s-Renyi graph, or simply E-R graph. The latter considers a graph with $N$ nodes and a probability $p$ to generate each edge, so that there are around $p \cdot \frac{N(N-1)}{2}$ edges, resulting in a binomial degree distribution:
\begin{equation} \label{eq:binomial}
P(k) = \binom{N - 1}{k}p^{k}(1 - p)^{n-1-k}
\end{equation} 
\noindent now, if $N \to \inf$ and $Np = const$, this degree distribution converges to a Poissonian distribution:
\begin{equation} \label{eq:poisson}
P(k) \sim e^{-pN} \cdot \frac{(pN)^{k}}{k!}
\end{equation} 
While this model represents an early attempt to describe real systems, a more advanced has been proposed by Albert and Barabasi, i.e. the BA model~\cite{barabasi01} that focuses on scale-free networks. This class of networks is characterized by a $P(k)$ that follows a power-law function as:
\begin{equation} \label{eq:scale_free}
P(k) \sim c \cdot k^{-\gamma}
\end{equation} 
\noindent with $c$ normalizing constant and $\gamma$ parameter of the distribution known as scaling parameter. 
In this class of networks, few nodes (called hubs) have many connections (i.e. a high degree), while the majority of nodes only few.
The BA model considers $N$ nodes and a second parameter, usually defined as $m$, representing the minimum number of edges per node.
In the thermodynamic limit, the BA model leads the network to be fitted by a power-law function with scaling parameter $\gamma = 3$, and to an average degree equal to $2m$.
Two main generative mechanisms allow to implement the BA model: the first-mover-wins and the fittest gets richer.
\subsection*{First-mover-wins}
The first-mover-wins constitutes the basic mechanism of the BA model. It can be summarized as follows:
\begin{enumerate}
\item Define $N$ number of nodes and $m$ minimum number of edges for each node
\item Add a new node and link it with other $m$ pre-exisisting nodes. Pre-existing nodes are selected according to the following equation:
\begin{equation} \label{eq:pref-att}
\Pi(k_{i}) = \frac{k_{i}}{\sum_{j} k_{j}}
\end{equation}
\noindent with $\Pi({k_{i}})$ probability that the new node generates a link with the $i$-th node having a $k_{i}$ degree. 
\end{enumerate}
This mechanism is not able to capture any difference among nodes.
\subsection*{Fittest gets richer}
Bianconi and Barabasi~\cite{bianconi01} proposed a variation to the previous mechanism by introducing a fitness parameter $\eta$. 
Here, the fitness parameter represents the ability of a node to compete for new links. In particular, the degree of the $i$-th node is proportional to:
\begin{equation} \label{eq:prob_link}
\Pi_{i} = \frac{\eta_{i} k_{i}}{\sum_{j} \eta_{j}k_{j}}
\end{equation}
\noindent with $k_{i}$ degree of the $i$th node. Notably, new nodes tend to link with pre-existing nodes having high values of $(\eta,k)$.
Therefore, even new nodes can reach a high degree if provided with a good fitness.
\subsection{Clustering Coefficient}
The clustering coefficient~\cite{barabasi01} allows to assess if nodes tend to cluster together. This phenomenon is common in many real networks as social networks, where it is possible to observe the emergence of circles of friends where people know each other. 
The clustering coefficient can be computed as:
\begin{equation}\label{eq:avg_cluster}
C = \frac{3 \times Tn}{Tp}
\end{equation}
\noindent $Tn$ is the number of triangles in the network and $Tp$ is the number of connected triples of nodes. A connected triple is a single node with edges running to an unordered pair of others. The value of $C$ lies in the range $0 \le C \le 1$.
A further way to compute this parameter has been proposed by Watts and Strogatz~\cite{watts01}, focusing on this quantity at a local level, i.e. computing the value
\begin{equation}\label{eq:avg_cluster_watts_local}
C_{i} = \frac{Tn_{i}}{Tp_{i}}
\end{equation}
\noindent with $Tn_{i}$ number of triangles connected to node $i$ and $Tp_{i}$ number of triples centered on node $i$. In this case, the local $C$ of nodes with a degree equal to $0$ or $1$ is set to $0$. In doing so, the global clustering coefficient of a network can be computed as
\begin{equation}\label{eq:avg_cluster_watts_global}
C = \frac{1}{n} \sum_{i} C_{i}
\end{equation}
This parameter allows to measure the density of triangles in a network, and can be computed both in directed and undirected networks. 
\subsection*{Path Length}
The distance between two nodes, belonging to the same network, can be computed considering the edges (and their weights in weighted networks) generating the path between them. A geodesic corresponds to the minimum path between two nodes, while a distance becomes infinite when there are no paths between them. The Dijkstra's algorithm~\cite{mohring01} and the Floyd-Warshall algorithm \cite{han01} are two classical algorithms for computing this parameter, however often the utilization of heuristics might be useful since this task can be computationally very expensive. Eventually, it is worth to highlight that the average shortest path length provides a first clue for assessing if a network is small-world~\cite{watts01}. Notably, 'small-world' networks have an average distance $L$ between pairs of nodes that scales with the size of the network, i.e. $L \propto \ln N$.
\section{Results}\label{sec:result}
\begin{figure}
%\centering
\includegraphics[width=2.5in]{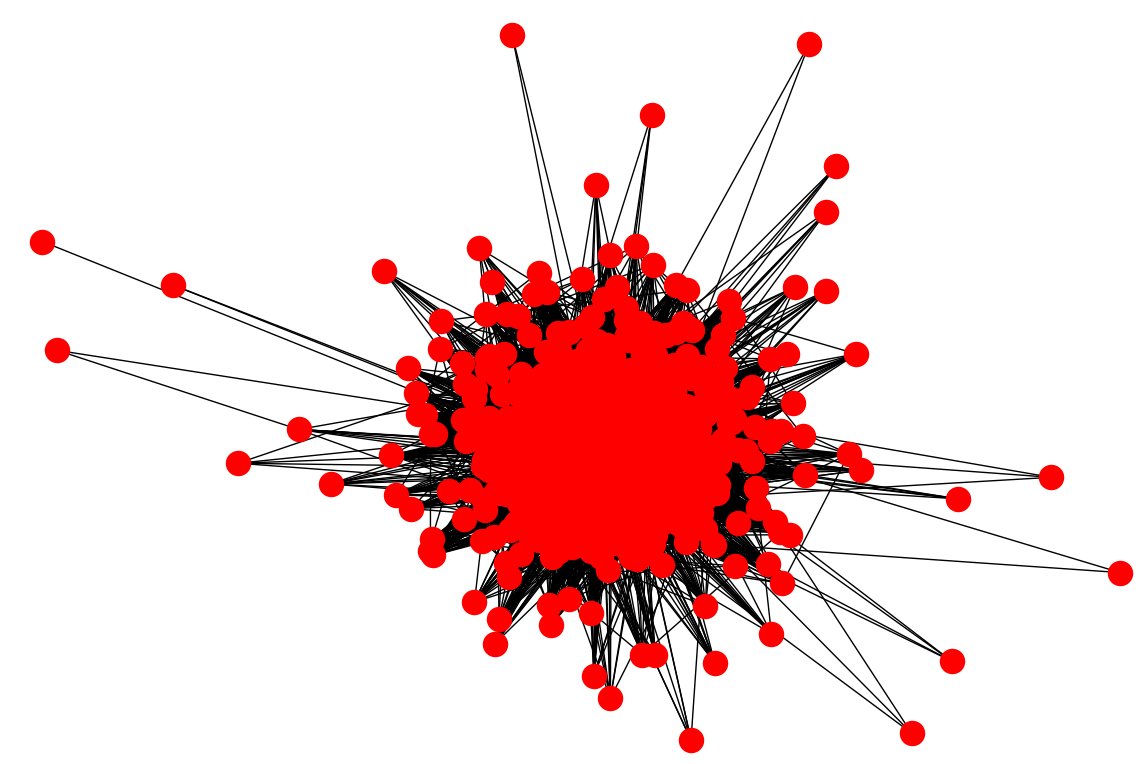}
\caption{Pictorial representation of the Bitcoin Cash Network related to the month of August 2017. A giant core strongly connected clearly emerges. Each link should be represent by an arrow, since our dataset represents a directed network, however here they are drawn as simple lines.}
\label{fig:figure_0}
\end{figure}

In this section, we describe results of our preliminary investigations on two kinds of networks, i.e. Bitcoin network and Bitcoin Cash network (see a pictorial representation in fig.~\ref{fig:figure_0}).
\subsection*{Dataset and Methodology}
The dataset related to Bitcoin network contains $7025$ nodes and refers to the month of April 2016, while those related to Bitcoin Cash refer to August and December 2017, and contain $963$ and $1454$ nodes, respectively.
The datasets used in this investigation refers to a subset of the related real networks. In particular, data have been obtained by using a cluster of machines connected to the Bitcoin (and Bitcoin Cash) network. Some of these nodes act as miners, while others perform only simple functions (e.g. routing).

The resulting cluster is able to obtain information about a large amount of other nodes according to the mechanism previously described (see Section~\ref{sec:bitcoin_network}), i.e. when a new node joins the network, it gets connected with some pre-existing nodes. Here, the communication protocol allows nodes to share the address of first order (i.e. nearest) neighbors and second order neighbors (i.e. friends of friends), then supporting the generation of new connections. 
Therefore, this mechanism has been exploited for achieving topological information and generating our datasets. So, while obviously the whole real networks are bigger than ours, we deem that the analyzed data provide a good description of the general behavior of the whole network.
\subsection*{Analysis}
\begin{figure}
%\centering
\includegraphics[width=2.5in]{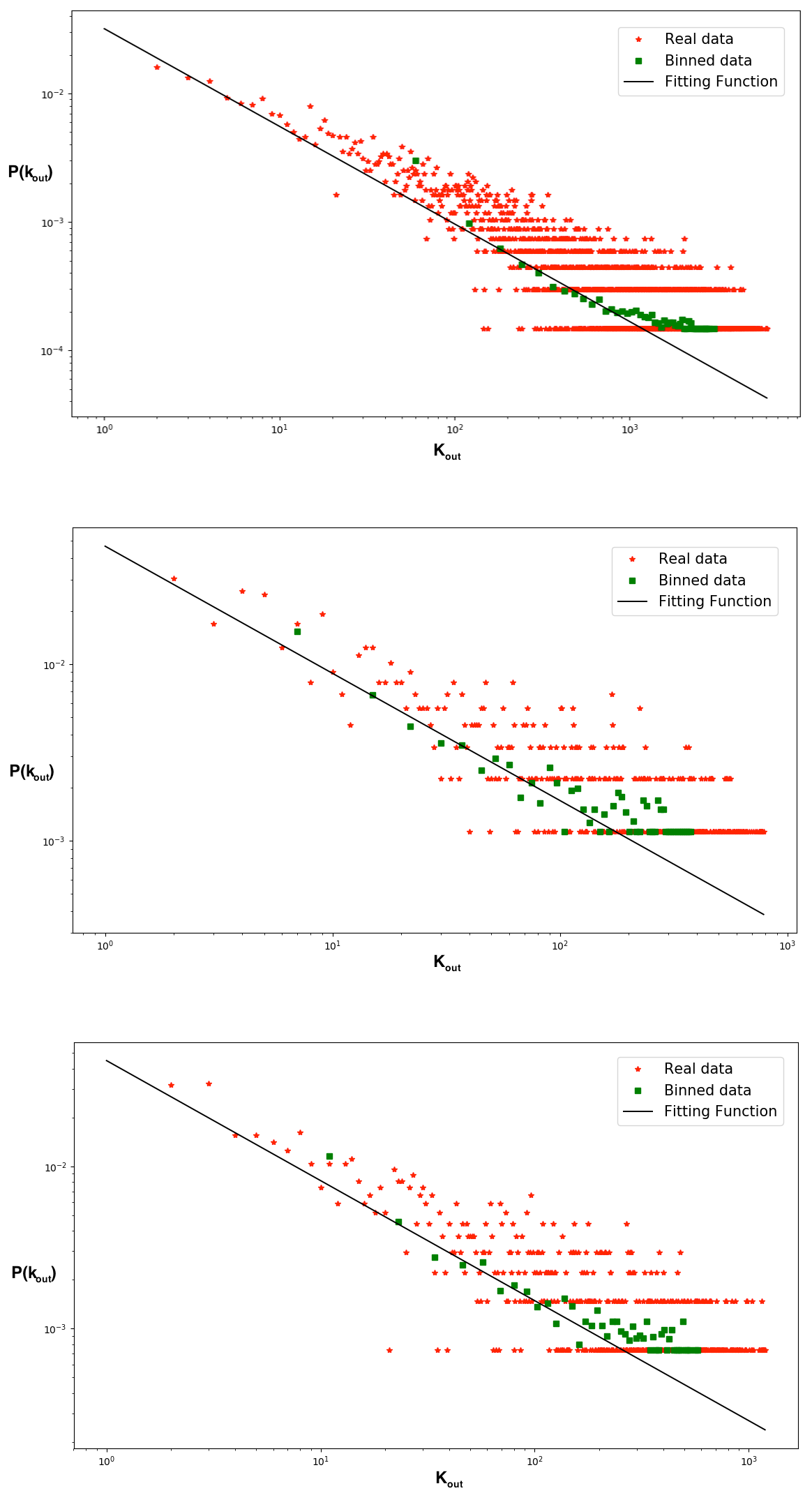}
\caption{Out-degree distributions $P(k_{out})$. From top to bottom: Bitcoin network on April 2016, Bitcoin Cash network on August 2017, and Bitcoin Cash Network on December 2017. As reported in the legend, red stars represent real data, green squares represent binned samples, and the black lines represent the power-law fitting curves.}
\label{fig:figure_1}
\end{figure}

\begin{figure}
\centering
\includegraphics[width=2.5in]{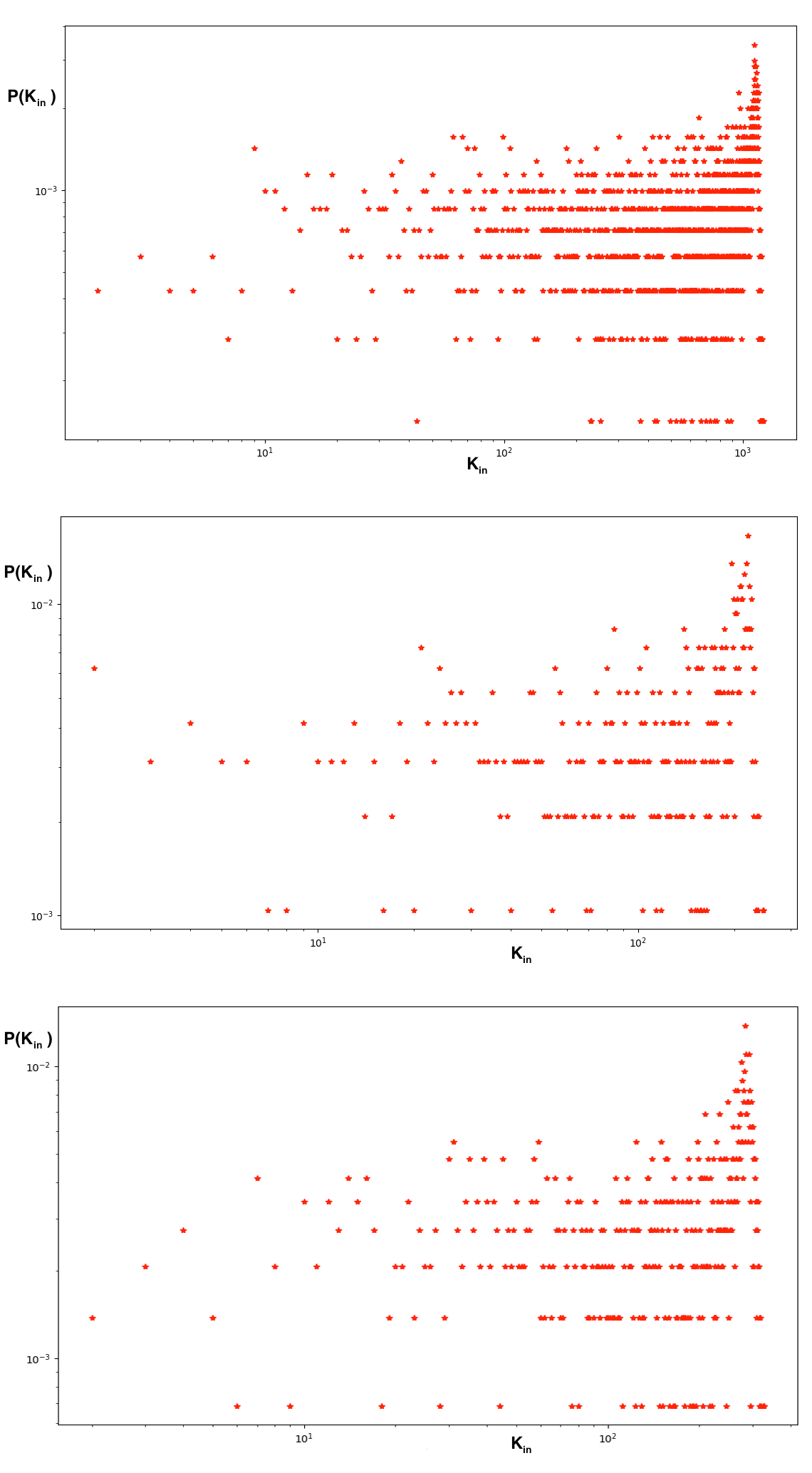}
\caption{In-degree distributions $P(k_{in})$. From top to bottom: Bitcoin network on April 2016, Bitcoin Cash network on August 2017, and Bitcoin Cash Network on December 2017.}
\label{fig:figure_2}
\end{figure}

According to their dynamics, these networks are directed, i.e. nodes are connected by links that can be represented as arrows. As preliminary analysis, we focus on the degree distribution. In particular, the out-degree distribution representing that of arrows leaving a node, and the in-degree distribution representing the distribution of arrows reaching a node.
Results related to the out-degree distribution (i.e. $P(k_{out})$) of the two Bitcoin Cash networks are shown in fig.~\ref{fig:figure_1}, while the in-degree (i.e. $P(k_{in})$) of the same networks are shown in fig.~\ref{fig:figure_2}.
In general, we found that few nodes have a minimum in-degree or out-degree equal to zero. So, we decided to remove them from the analysis, considering a minimum degree (for both cases) equal to $1$.
Considering the resulting out-degree distributions, we hypothesized that they might be fitted by a power-law equation. Therefore, we computed the related scaling parameters (i.e. $\gamma_{out}$). As widely discussed in~\cite{clauset01}, when the minimum degree is $k_{min} = 1$, the computing of scaling parameters requires particular attention (see also~\cite{goldstein01}).
Notably, a possible approach is based on the maximum likelihood estimator (MLE), that requires to find the value of $\gamma$ (in our case $\gamma_{out}$, since it refers to the out-degree distribution) that satisfies the following equation:

\begin{equation}\label{eq:gamma_mle}
\frac{\dot{\zeta(\gamma)}}{\zeta(\gamma)} = \frac{1}{n} \sum_{i=1}^{n} \log(k_i)
\end{equation}

\noindent with $\zeta$ Riemann Zeta function and $\dot{\zeta}$ its derivative. Some numerical values of the ratio $\frac{\dot{\zeta(\gamma)}}{\zeta(\gamma)}$ are available in~\cite{walter01}.
Following this method, we achieved for all three networks values of $\gamma_{out}$ smaller than $1$ (see Table~\ref{tab:statistical}).
These results indicate that both kinds of network (i.e. Bitcoin and Bitcoin Cash) have an out-degree distribution that might be fitted by a power-law, however further analyses for the goodness-of-fit are required. Instead, the in-degree distributions seem to be more homogeneous, since in general all nodes have a similar in-degree.
Then, we analyze the average clustering coefficient (i.e. $C$) and the average shortest path length (i.e. $SPL$) of the three networks. These two parameters have been computed using the Python library NetworkX~\cite{networkx01}.
Table~\ref{tab:statistical} reports the related numerical values.
\begin{center}
{\footnotesize\sc Network Parameters}
\vskip .9\baselineskip
{\def\arraystretch{1.5}\tabcolsep 5pt\footnotesize
\begin{tabular}{lllll}
\hline
Network & $N$ & $\gamma_{out}$  & C & SPL\\
\hline
BC Apr 2016& 7025 & 0.76 & 0.486 & 1.816\\
BCH Aug 2017& 963& 0.72 &0.576 & 1.7267\\
BCH Dec 2017& 1454& 0.74 &0.544 & 1.7605\\
\hline
\end{tabular}}\label{tab:statistical}
\end{center}
It is worth to highlight that all networks seem to be provided with a small-world behavior, since their $SPL$ is smaller than the logarithm of their size (i.e. $N$).

\section{Discussion and Conclusion}\label{sec:conclusion}
In this work, we perform a preliminary analysis of a real Bitcoin Network and two Bitcoin Cash networks.
Our aim is both analyzing their connectivity pattern and trying to understand 'how' their structure evolved. 
In order to reach this goal, we need to remind how a Bitcoin network works, as described in Section~\ref{sec:bitcoin_network}. In particular, as we know, since in principle all nodes have to store the whole 'ledger' of transactions, the in-degree distribution, i.e. the information received, must be more or less uniform (remarkably an homogeneous distribution actually characterizes their structure).
On the other hand, since storing connections requires resources (e.g. memory), in general most nodes remain connected to few neighbors. At the same time, powerful nodes can store more node connections than others. Thus, considering the generative mechanisms before described, i.e. 'fittest-gets-richer' and 'first-mover-wins', we think that the former might be more suitable for modeling the evolution of the out-degree distribution. In particular, the fitness can be related to the computational power of nodes, i.e. the higher the fitness the higher the availability of resources (and then the out-degree).
In addition, even in the light of the increasing amount of Raspberry Pi, appearing in the network, we suggest that a 'fitness based' model can be effective in describing the dynamics of the Bitcoin network and the Bitcoin Cash network.
Moreover, since the structure of the networks (considering their out-degree distribution), seems very similar to that of scale-free networks, we have a first clue that they can be 'small-world'~\cite{watts01}. Here, a small-world behavior would be essential for the reliability of these networks. 
Therefore, we analyzed two important parameters, i.e. the average clustering coefficient and the average shortest path length. While the latter suggests that all these three networks are 'small-world', the former can be used also for future investigations. Notably, we aim to compare the average clustering coefficient with that achieved in an E-R network generated with the same number of nodes and having a similar number or edges (this result can be achieved by setting an opportune value for the parameter related to the edge probability before described). 
In particular, this analysis might constitute a further proof, combined with the average shortest path length, of the small-world behavior of a network.
Finally, it is important to observe that beyond the difference in the size (i.e. in the amount of nodes), the Bitcoin network and the two Bitcoin Cash networks appear very similar. In general, that difference can be expected, being the Bitcoin Cash network recently proposed. 
To conclude, beyond the further analyses above mentioned, as for future work we deem relevant to investigate also the evolution of stochastic processes on these structures. In particular, analyzing spreading phenomena and percolation, can be useful for a better understanding on how these modern technologies behave sharing data and information.

%
% The following two commands are all you need in the
% initial runs of your .tex file to
% produce the bibliography for the citations in your paper.
%\bibliographystyle{abbrv}
%\bibliography{sigproc}  % sigproc.bib is the name of the Bibliography in this case

\begin{thebibliography}{99}
%
\bibitem{parallel01}
Keckler, S.W., et al.:
GPUs and the future of parallel computing.
\emph{IEEE Micro } \textbf{31-5} 7--17 (2011)

\bibitem{satoshi01}
Satoshi, N.:
Bitcoin: A Peer-to-Peer Electronic Cash System.
\emph{https://bitcoin.org/bitcoin.pdf} (2009)

\bibitem{block01}
Crosby, M., et al.:
Blockchain technology: Beyond bitcoin.
\emph{Applied Innovation Review } \textbf{2} (2016)

\bibitem{barronchelli01}
ElBahrawy, A., et al.:
Evolutionary dynamics of the cryptocurrency market.
\emph{R. Soc. open sci.} \textbf{4}, 170623 (2017)

\bibitem{barabasi01}
Albert,R. and Barabasi, A.L.:
Statistical Mechanics of Complex Networks.
\emph{Rev. Mod. Phys} \textbf{74}, 47--97 (2002)

\bibitem{bianconi01}
Bianconi, G. and Barabasi, A.L.:
Bose-Einstein condensation in complex networks.
\emph{Physical Review Letters} \textbf{86}, 5632 (2001)

\bibitem{javarone01}
Javarone, M.A., Armano, G.:
Quantum Classical Transition in Complex Networks.
\emph{Journal of Statistical Mechanics: Theory and Experiment} \textbf{4}, P04019 (2013)

\bibitem{antinopolis01}
Antonopoulos, A.M.:
Mastering Bitcoin: Unlocking Digital Cryptocurrencies.
\emph{OReilly} (2014)

\bibitem{estrada01}
Estrada, E.:
The Structure of Complex Networks: Theory and Applications.
\emph{Oxford University Press} (2011)

\bibitem{erdos01}
P. Erdos, P. and Renyi, A.:
On the Evolution of Random Graphs.
\emph{pubblication of the mathematical institute of the hungarian academy of sciences} 17--61 (1960)

\bibitem{watts01}
Watts, D. J. and Strogatz, S. H.:
Collective dynamics of ``small-world'' networks.
\emph{Nature} 440--442 (1998)

\bibitem{mohring01}
Mohring, R.H., Schilling, H., Schutz, B., Wagner, D., Willhalm, T.:
Partitioning Graphs to Speed Up Dijkstra's Algorithm.
Experimental and Efficient Algorithms LNCS, 3503, 189--202 (2005)

\bibitem{han01}
Han, S.C., Franchetti, F., Pushel, M.:
Program Generation for all-pairs shortest path problem.
Proc. of the 15th Int.Conf. on Parallel architectures and compilation techniques, 22--232 (2006)

\bibitem{clauset01}
Clauset, A., Shalizi, C.R., Newman, M.:
Power-Law Distributions in Empirical Data.
\emph{SIAM} \textbf{51(4)} 661--703 (2009)

\bibitem{goldstein01}
Goldstein, M.L., Morris, S.A., Yen, G.G.:
Problems with fitting the power-law distribution.
\emph{EPJ-B} \textbf{41(2)} 255--258 (2004)

\bibitem{walter01}
Walther, A.:
Anschauliches zur Riemannschen Zetafunktion.
\emph{Acta Math} \textbf{3-4(48)} 393--400 (1926)

\bibitem{networkx01}
Hagberg, A.A., Schult, D.A., Swart, J.:
Exploring network structure, dynamics, and function using NetworkX.
\emph{Proceeding of th 7th Python in Science Conference} 11--15 (2008)

\end{thebibliography}
% You must have a proper ".bib" file
%  and remember to run:
% latex bibtex latex latex
% to resolve all references
%
% ACM needs 'a single self-contained file'!
%
%APPENDICES are optional
%\balancecolumns

\end{document}